\documentclass[12pt]{article}

\usepackage{amsmath}
\usepackage{graphicx}
\usepackage{amssymb}
\usepackage{amsfonts}
\usepackage{color}
\usepackage{indentfirst}
\usepackage{latexsym}

\def\beq{\begin{eqnarray}}
\def\eeq{\end{eqnarray}}
\def\nn{\nonumber}                

\def\ln{\,\mbox{ln}\,}

\def\tr{\,\mbox{tr}\,}

\def\Tr{\,\mbox{Tr}\,}

\def\al{\alpha}
\def\be{\beta}

\def\ga{\gamma}
\def\de{\delta}
\def\vp{\varepsilon}

\def\ze{\zeta}

\def\la{\lambda}
\def\na{\nabla}
\def\pa{\partial}

\def\si{\sigma}

\def\ph{\varphi}

\def\Ga{\Gamma}
\def\De{\Delta}

\DeclareMathOperator{\cx}{\square}

\begin{document}

\begin{center}
\renewcommand*{\thefootnote}{\fnsymbol{footnote}}

{\Large
Anomaly-induced vacuum effective action with torsion:
\\
covariant solution and ambiguities}
\vskip 6mm
		
{Guilherme H. S. Camargo}
\hspace{-1mm}\footnote{E-mail address: \ guilhermehenrique@unifei.edu.br}
\ \ and \ \ {Ilya L. Shapiro}
\hspace{-1mm}\footnote{E-mail address: \ ilyashapiro2003@ufjf.br}
\vskip 6mm
		
{\sl
Departamento de F\'{\i}sica, ICE, Universidade Federal de Juiz de Fora,
\\
Juiz de Fora, 36036-900, Minas Gerais, Brazil}
\end{center}
\vskip 2mm
\vskip 2mm
	
	
\begin{abstract}
		
\noindent
The conformal anomaly in curved spacetime with antisymmetric
torsion is reconsidered,  taking into account new important details.
We formulate, for the first time, the covariant solution of the
anomaly-induced effective action.
The covariant effective action includes local terms corresponding
to total derivatives in the conformal anomaly. The contribution of
massless fermions to these terms is characterized by multiplicative
anomaly, coming from two different choices of doubling for the
spinor operator. On the other hand, the nonlocal part of
anomaly-induced action is free of ambiguities and admits a
low-energy limit, similar to the effective potential in the
metric-scalar theory.
\vskip 4mm
		
		
\end{abstract}
	
\setcounter{footnote}{0} 
\renewcommand*{\thefootnote}{\arabic{footnote}} 
\section{Introduction}
\label{sec1}

General features of conformal anomaly and the induced
action of gravity corresponding to anomaly
represent an interesting and active subject
of interest starting from the epoch when anomaly was discovered
\cite{CapDuf-74,duff77,ddi,duff94} and anomaly-induced
effective action derived in two \cite{polyakov81} and four ($4D$)
\cite{rie,frts84} spacetime dimensions. The reason for this special
interest is related to important applications to black hole physics
\cite{ChrFull,balsan}, cosmology \cite{fhh,star,StabInstab} (see
also \cite{duff94,PoImpo,OUP,Mottola-SM} for review and further
references), effective
approaches to quantum gravity based on the anomaly-induced action
in $4D$ \cite{odsh90-CQG,AntMot92,induce} and possible
nonperturbative generalizations in the form of $a$- and $c$-theorems
(see, e.g., \cite{KamSch-2011,LPR-2012,Kamarg-2012}).

From a more general perspective, the anomaly is technically simple
and elegant way for describing loop corrections in the semiclassical
approach. Usually, the trace anomaly is associated with the UV
limit because classical conformal symmetry is typical only in a
massless theory. In other words, the anomaly-induced action is a
direct generalization of the renormalization group improved classical
action based on the Minimal Subtraction scheme of renormalization.
The generalization, in this case, means that the constant rescaling of
the metric (curved-space equivalent of rescaling the momenta
\cite{nelspan82,tmf,Toms83}) is replaces by the local conformal
transformation with the conformal factor depending on the
coordinates. On the other hand, the anomaly-induced action can be
adapted to describe the low-energy (IR) limit of a massless theory.
The first examples of this sort can be found in \cite{MotVaul} and
\cite{AnoScal}, where the anomaly has been used to obtain the
one-loop effective potential of scalar fields in curved spacetime.

Anomaly is directly related to the logarithmic divergences in the
conformal massless theory \cite{duff94,OUP}. This is true for a
purely metric background, but also for the theory with extra
background fields, such as scalars, vector fields, and torsion.
One can prove that the divergences satisfy the conformal Noether
identity \cite{tmf} and, for this reason, the anomaly is composed
by the following three types of terms \cite{ddi,DeserSchwimmer}:

\textit{i)} Topological term, such as the Gauss-Bonnet term in
$4D$, or its analogs in higher even dimensions. The conformal
transformation of this term greatly simplifies if supplemented by
a specially chosen surface terms. This fundamental feature is the
main basis of integrating anomaly, which was verified in $2D$
\cite{polyakov81}, in $4D$ \cite{rie,frts84} and in $6D$ \cite{6d}.
The general proof for higher dimensions is not known and remains
a conjecture \cite{FST}. Assuming this is correct, the topological
term is the main source of the nonlocal structures in the
anomaly-induced action.

\textit{ii)} Legitimate conformal terms (we will call them
$C$-terms), such as the square of the Weyl tensor in $4D$; three
possible conformal structures in $6D$ \cite{BFT-2000}, etc. The
integration of these terms is relatively simple assuming the
aforementioned conjecture. The result for the covariant induced
action is nonlocal.

\textit{iii)} The total derivative terms in the divergences provide
the same terms in the anomaly. These terms are known to be
ambiguous \cite{duff94} and this ambiguity equivalent to adding
a finite local covariant nonconformal term to the classical action,
as discussed in \cite{anomaly-2004,BoxAno} for dimensional
and Pauli-Villars regularizations.
Strictly speaking, \textit{i)}- and \textit{iii)}-type terms are not
conformal invariant. However, those are surface terms that satisfy
the conformal Noether identity. For this reason, we shall call them
$N$-terms.

Verifying the generality of the described classification may go
in two different ways, namely either
increasing the dimension or trying to enrich space-time geometry.
The first approach is extremely difficult for practical realization
(see, e.g., \cite{6d}). At the same time, the second possibility
may be related to introducing torsion or nonmetricity of the
spacetime. The present work reports on the verification of the
scheme described above for the theory with torsion. Namely,
we explore the uniqueness of the topological term, the
possibility to construct covariant forms for anomaly-induced
action with torsion, and the ambiguities in the local covariant
terms, responsible for the total derivatives in the anomaly.

The case of the anomaly with torsion has been explored in several
works \cite{Obukhov83,buodsh85}, including for the different
realizations of conformal symmetry \cite{anhesh,torsi}. However,
the covariant (nonlocal and local) form of the anomaly-induced
effective action of gravity with torsion was never formulated and
the first purpose of the present work is to fill this gap. We shall
see that the integration of all terms of an anomaly in the theory
with torsion can be done in a very standard way, with one
important addition. As we are working with fermions, there is
always a possibility to have an ambiguity related to different
ways of doubling the Dirac operator. Previously, it was shown
that, in case of massive fermions, this ambiguity leads to the
nonlocal multiplicative anomaly \cite{QED-Form},
something one cannot consistently achieve \cite{Evans,Dowker}
using $\ze$-regularization \cite{zeta} owing to the presence of
renormalization $\mu$-dependence in the local terms in the
effective action. In what follows we show that, in the theory with
torsion, the multiplicative anomaly is possible even in the
\textit{massless} case, i.e., in the local terms coming from the
integration of total derivative terms in the trace anomaly. Since
these terms are not renormalized, this new version of multiplicative
anomaly avoids the $\mu$-dependence and the arguments of
\cite{Evans,Dowker}.

The paper is organized as follows. In Sec.~\ref{sec2} we briefly
review the notations for gravity with torsion and define the actions
of free matter fields with conformal symmetry. Sec.~\ref{sec3}
describes the calculation of one-loop divergences in the fermion
case, generalizing the previous works on the subject
\cite{Obukhov83,buodsh85,anhesh,torsi}.
Since one of our main concerns is ambiguity in the anomaly, we
perform the calculation in two different ways and meet the
difference which leads to the local multiplicative anomaly in
the effective action.
In Sec.~\ref{sec4} the anomaly is used to find the covariant
(nonlocal and local) solutions for the anomaly-induced vacuum
effective action. Sec.~\ref{sec5} describes the low-energy limit
in the effective action in the metric-torsion theory, constructed in
analogy to the effective potential of a scalar field \cite{AnoScal}.
Finally, in Sec.~\ref{sec6} we draw our conclusions.

\section{Conformal fields with torsion}
\label{sec2}

In what follows we shall give only a brief list of necessary formulas
about gravity with torsion and conformal matter fields. The path
integrals over these fields require renormalization of vacuum
action and produce trace (conformal) anomaly. A more detailed
review can be found, e.g., in \cite{Hehl-76,torsi}. In the last
reference, the notations are the same as here.

The affine connection without torsion is the Christoffel symbol
(Levi-Civita connection)
\beq
\Gamma^\alpha_{\,\beta\gamma}
 = \big\{^{\,\al}_{\be\ga}\big\} =
\frac12\,g^{\al\la}\,\big(
\pa_\be\,g_{\la\ga} + \pa_\ga\,g_{\la\be} - \pa_\la\,g_{\be\ga} \big).
\label{1.5}
\eeq
The corresponding covariant derivative satisfies the metricity
condition $\na_\la g_{\al\be} = 0$ and is free of torsion, i.e.,
$\Ga^\tau_{\,\al\be} = \Ga^\tau_{\,\be\al}$. In what follows, we do
not consider the theories with nonmetricity, but include nonzero
torsion, that is making geometry more extensive and, in particular,
links it to the spin of matter fields \cite{Hehl-76}.

Torsion tensor is defined as the difference between the two affine
connections which are not assumed symmetric,
\beq
T^\alpha_{\,\cdot\,\beta\gamma}
= \tilde{\Gamma}^\alpha_{\,\beta\gamma}
- \tilde{\Gamma}^\alpha_{\,\gamma\beta}.
\label{torsion}
\eeq
It is useful to present torsion tensor as a sum of the irreducible
components \cite{bush90} (see also \cite{torsi}),
\beq
T_{\alpha\beta\mu} = \frac{1}{3}
\left(T_{\beta}\,g_{\alpha\mu} - T_{\mu}\,g_{\alpha\beta}\right)
- \frac{1}{6}\, \varepsilon_{\alpha\beta\mu\nu}\,S^{\nu}
+ q_{\alpha\beta\mu}\,,
\label{irr}
\eeq
namely the vector $T_\be = T^\al_{\,\cdot\,\be\al}$, \ axial vector
$S^{\nu} = \epsilon^{\alpha\beta\mu\nu}T_{\alpha\beta\mu}$,\
and the remaining tensor
$\;q^\alpha_{\,\cdot\,\beta\gamma}$,
satisfying the conditions $q^\alpha_{\;\cdot\,\beta\alpha} = 0$
and $\epsilon^{\alpha\beta\mu\nu}q_{\alpha\beta\mu} =0$.

One can derive the generalizations of the Riemann tensor, Ricci
tensor and scalar curvature for the connection with torsion, e.g.,
\beq
 {\tilde R}
 \,=\,
 R - 2\na_\al T^\al - \frac{4}{3} T_\al T^\al
+ \frac{1}{2} q_{\alpha \beta \gamma} q^{\alpha \beta \gamma}
+ \frac{1}{24} S^\al S_\al\,.
\label{irred}
\eeq
Since our purpose is to consider the quantum theory of matter
fields, regarding both metric and torsion as external fields, the
parametrization of these background fields is mostly irrelevant.
Thus, in what follows, we shall use the Riemannian version of
the curvature tensors.

Since the interaction of torsion with the gauge fields is forbidden
by the gauge invariance \cite{Hehl-76}, we shall assume that gauge
vectors decouple from torsion at the classical level. The same
symmetry protects the theory from these interactions at the
quantum level \cite{bush85,torsi}. Thus, we need to consider only
scalar and fermion fields, as described below.

\subsection{Scalar field}
\label{sec2.1}

The action of the real nonminimal scalar field $\phi$ in curved
spacetime with torsion has the form
\beq
S_0 \,=\,\frac12 \int d^4x \sqrt{-g}\,\big\{
g^{\al\be}\pa_\al \phi\pa_\be \phi
+ \xi_i P_i \phi^2
- m^2\phi^2 \big\},
\label{scalaction}
\eeq
where
the nonminimal parameters $\xi_i$ correspond to the
structures
\beq
P_1 = R,
\quad
P_2 = \na_\al T^\al,
\quad
P_3 = T_\al T^\al,
\quad
P_4 = q_{\alpha \beta \tau} q^{\alpha \beta \tau},
\quad
P_5 = S_\al S^\al,
\label{nonmi}
\eeq
repeating the ones of (\ref{irred}) but with arbitrary coefficients
 $\xi_i$.
It is known \cite{bush85,bush90,torsi} that in the curved-space
theory with fermions, scalars and Yukawa coupling, arbitrary
nonminimal parameters $\xi_1$ and  $\xi_5$, of the interaction
of scalar field(s) with $R$ and $S_\al$, are needed to provide
renormalizable semiclassical theory, as will be explained below.
In this article, we will mainly restrict the consideration
by a purely antisymmetric
torsion. The reasons are that  \ \textit{i)} this is the most
relevant part of the matter-torsion interaction, in particular
linking spin to geometry \cite{Hehl-76};
\ \textit{ii)} more general cases are not expected to bring new
details, concerning the aforementioned aspects of the anomaly
and also compared to the previous analysis in \cite{torsi}.
Thus, we shall assume that $T_\mu=0$ and $q_{\al\be\tau}=0$,
such that only the axial vector component in (\ref{irr}) is present.

Action (\ref{scalaction}) is invariant under general coordinate
transformations. On top of that, the massless model with $\xi=1/6$
is invariant under the transformation called local conformal
symmetry,
\beq
g_{\mu\nu}=e^{2\si}\,{\bar g}_{\mu\nu},
\qquad
S_\al = {\bar S}_\al,
\qquad
\phi=e^{-\si}\,{\bar \phi},
\label{confscal}
\eeq
where $\si=\si(x)$. Note that the value of
$\xi_5$ does not affect conformal invariance. Also, if we include
$T_\mu$ component, there are three types of the conformal
transformations with torsion. This subject was discussed in detail
in \cite{bush85,anhesh} and we will not repeat it here.

\subsection{Massless Dirac field}
\label{sec2.2}

The interaction of Dirac spinor field with torsion is described
by the parity-preserving action
\beq
S \,=\, \int d^4x \sqrt{-g}\,\bar{\psi} \Big\{
i\ga^\mu \big(\na_\mu - i\eta \ga^5 S_\mu - i \eta_2 T_\mu\big)
- m \Big\} \psi
\label{S0}
\eeq
with the nonminimal parameters $\eta$ and $\eta_2$. It is important
that the minimal coupling of fermion with torsion correspond to
$\eta=1/8$ and $\eta_2=0$ \cite{Hehl-76}. This feature explains the
difference between interaction of torsion with $S_\mu$
and $T_\mu$. Starting from the minimal actions, there is no $T_\mu$
and $q_{\alpha\beta\mu}$, and the corresponding interactions never
emerge in the divergences. On the contrary, $S_\mu$ is always present
in both classical theory and in the divergences. As a result, one has
to renormalize the parameters $\eta$ and also $\xi_5$, in case the
theory includes Yukawa interactions between scalars and fermions.
Thus, one cannot have a renormalizable theory based on the minimal
coupling to the external torsion.

In the massless case, the theory (\ref{S0}) possesses three different
symmetries. One of those is the usual Abelian gauge symmetry related
to $T_\mu$. In fact, one can trade $\eta_2 T_\mu$ by $e A_\mu$ and,
in the part of the gauge symmetry, reduce the problem to the usual
gauge field. Since we are interested in the $T_\mu=0$ case, the
corresponding transformation will not be considered. Another
symmetry is
\beq
\bar{\psi} \,\longrightarrow\,\bar{\psi}e^{i\eta \al \ga^5},
\qquad
\psi \,\longrightarrow\,e^{i\eta \al \ga^5}\psi,
\qquad
S_\mu \,\longrightarrow\,S_\mu + \pa_\mu \al,
\label{chiral}
\eeq
where $\al=\al(x)$ is a scalar transformation parameter.
Finally, there is a conformal transformation of the spinor field,
supplementing the one of  (\ref{confscal}),
\beq
g_{\mu\nu}=e^{2\si}\,{\bar g}_{\mu\nu},
\qquad
S_\al = {\bar S}_\al,
\qquad
\bar{\psi} = \bar{\psi}e^{-\frac32\,\si},
\qquad
\psi = \psi e^{-\frac32\,\si}.
\label{confspin}
\eeq
Let us mention that the values of torsion nonminimal parameters,
$\eta$ and $\xi_5$, do not affect the conformal invariance.

The rest of this work is devoted to the anomaly in the local
conformal symmetry (\ref{confscal}), (\ref{confspin}) in the
vacuum part of the one-loop effective action $\Ga^{(1)}(g,S)$.

According to the general proof \cite{tmf}, if the classical
actions of quantum matter fields have local conformal symmetry
(\ref{confscal}) and  (\ref{confspin}), the divergent part
$\Ga_{div}^{(1)}(g,S)$ satisfies the corresponding Noether identity,
\beq
-\,\frac{2}{\sqrt{-g}}\,g_{\al\be}
\,\frac{\de \Ga_{div}^{(1)}(g,S)}{\de g_{\al\be}}
\,=\,
\Phi(g,S) .
\label{confNo}
\eeq
Here $\Phi(g,S)$ is a covariant finite expression, that is also local
owing to the Weinberg's theorem. We note that Eq.~(\ref{confNo})
does not include the variational derivative with respect to $S_\mu$
because, according to (\ref{confscal}), its conformal weight is zero.

The possible vacuum divergences obey the mentioned
symmetries. The Riemannian terms include the square of the
Weyl tensor in $4D$,
\beq
C^2\,=\,
R_{\al\be\mu\nu}R^{\al\be\mu\nu}
- 2 R_{\al\be}R^{\al\be} + \frac13 R^2,
\label{Weyl-1}
\eeq
the integrand of the Gauss-Bonnet topological term
\beq
E_4\,=\,R_{\al\be\mu\nu}R^{\al\be\mu\nu}
- 4 R_{\al\be}R^{\al\be} + R^2,
\label{GB}
\eeq
and the surface term $\cx R$. The important difference between
these terms is that, in $4D$, the integral of $C^2$ is invariant,
\beq
\int d^4x \sqrt{-g}\, C^2
\,=\,
\int d^4x \sqrt{-\bar{g}}\,\bar{C}^2.
\label{C}
\eeq
We shall call the actions satisfying this condition
$\mathcal{C}$-terms. On
the contrary, the integrals $\int d^4x \sqrt{-g} E_4$ and
$\int d^4x \sqrt{-g} \cx R$ do not possess this property and, strictly
speaking, are not conformal invariant. At the same time, both terms
satisfy the conformal Noether identity. All such actions, that are not
really conformal, but obey the rule
\beq
-\,\frac{2}{\sqrt{-g}}\,g_{\al\be}
\,\frac{\de S(g,\,S}{\de g_{\al\be}}\,=\,0,
\label{N}
\eeq
will be called $\mathcal{N}$-terms.

Besides the mentioned metric-dependent integrals, there are
torsion-dependent $\mathcal{C}$- and $\mathcal{N}$-terms. In the first group, there
are two new candidates \cite{book}
\beq
S^4 \,=\,(S^2)^2\,=\,\big(S_\mu S^\mu\big)^2
\qquad
\mbox{and}
\qquad
S_{\mu\nu}^2 \,=\,g^{\mu\al}g^{\nu\be}S_{\mu\nu}S_{\al\be},
\label{CS}
\eeq
where $\,S_{\mu\nu} = \na_\mu S_\nu - \na_\nu S_\mu$.
One can ask the following questions: \ \textit{i)} Whether there are
torsion-dependent analogs of (or alternatives to) the Riemannian
topological term $E_4$? \ \textit{ii)} Are there torsion-dependent
$\mathcal{N}$-terms, including total derivatives, similar to $\cx R$, and \ \
\textit{iii)} Whether the renormalization of these torsion-dependent
$\mathcal{N}$-terms has ambiguities (see \cite{duff94} and
\cite{anomaly-2004}) which are present in the cases of
$\cx R$ and, also, for a non-zero background scalar field, in the
$\cx \phi^2$-term \cite{BoxAno,AnoScal}. We shall address these
questions by making direct calculations of divergences, anomaly
and anomaly-induced action of gravity with torsion.

\section{Derivation of one-loop divergences}
\label{sec3}

For the sake of generality, we perform calculations for massive
versions of scalar and spinor fields and for an arbitrary $\xi_1$.
We can set masses to zero and $\xi_1=1/6$ at the end.

Let us start by quoting the known result for the scalar field
\cite{torsi}
\beq
&&
\Ga^{(1)}_{div,\, scal}
\,=\,
-\,\frac{\mu^{n-4}}{\vp}\,\int d^nx\sqrt{-g}\,\Big[
  \frac{1}{120}C^2
- \frac{1}{360}E_4
+ \frac{1}{180} \cx R
+ \frac16 \cx P
+ \frac12\,P^2
\Big],
\qquad
\qquad
\label{scalardivs}
\eeq
where
\beq
P = \Big(\frac16 - \xi_1\Big)R - \xi_5 S^2 + m^2.
\label{Pscal}
\eeq
In the massless limit and assuming $\xi_1=1/6$, we meet the first
$\mathcal{C}$-term from (\ref{CS}) and the $\mathcal{N}$-terms
i.e., $E_4$, $\cx R$, and $\cx  S^2$.

Consider the fermion contribution.
In this part, we go into the detail of the calculation,
regardless they can be partially found in \cite{torsi}. Our
purpose is to evaluate $-i \Tr \ln {\hat H}$, where
\beq
{\hat H}
\,=\,i\ga^\mu \big(\na_\mu - i\eta \ga^5 S_\mu\big) - m.
\label{H}
\eeq
In order to perform this calculation, we have to multiply (\ref{H})
by a conjugate operator, such that the product belongs to the
standard class of minimal operators
${\hat F}={\hat \cx} + 2{\hat h}^\al \na_\al + {\hat \Pi}$, admitting
application of the Schwinger-DeWitt technique \cite{DeWitt65,bavi85}.
It seems that there should be many possible choices for such a
conjugate operator, but there are two constraints. First of all,
the contribution of the conjugate operator should be calculable.
And, on the other hand, we have to respect the chiral symmetry
(\ref{chiral}), as otherwise the result may be wrong. The last means
the structure $\ga^\mu \big(\na_\mu - i\eta \ga^5 S_\mu\big)$ has to
be part of the conjugate operator or, alternatively, the conjugate
operator must be $S_\mu$-independent. In what follows, we consider
both these options and explore the difference.

\subsection{First calculation of fermion contributions}
\label{sec3.1}

As a first option, consider the conjugate operator of the
form
\beq
{\hat H}_1
\,=\,i\ga^\nu \big(\na_\nu - i\eta \ga^5 S_\nu\big) + m.
\label{H1}
\eeq
It is known that the change of the sign of the mass does not
change the result (see, e.g., \cite{GBP}), such that
$\Tr \ln {\hat H} = \Tr \ln {\hat H}_1$ and we can use
the relations 
\beq
- i \Tr \ln {\hat H}
\,=\,
- \frac{i}{2} \Tr \ln \big({\hat H} {\hat H}_1\big)
\,=\,
- \frac{i}{2} \Tr \ln
\big({\hat \cx} + 2{\hat h}_1^\al \na_\al + {\hat \Pi}_1\big).
\label{product1}
\eeq
After some algebra, we get
\beq
&&
{\hat h}_1^\al \,=\, \frac{i}{2}\,\ga^5
\big( \ga^\la \ga^\al  - \ga^\al \ga^\la \big)S_\la \,,
\nn
\\
&&
{\hat \Pi}_1 \,=\, m^2 - \frac14R + S^2
- i\ga^5 (\na_\al S^\al)  - \frac{i}{2}\ga^5 \ga^\al \ga^\be
S_{\al\be}.
\label{hPi1}
\eeq
In these and subsequent formulas we made a rescaling of the external
torsion field $\eta S_\mu \to  S_\mu$,  making formulas more compact.

The elements of the Schwinger-DeWitt technique are
 \beq
&&
{\hat P}_1
\,=\,
{\hat \Pi}_1 + \frac{{\hat 1}}{6}R
- \na_\al {\hat h}_1^\al -  {\hat h}_{1\al} {\hat h}_1^\al
\,=\, m^2 - \frac{1}{12}R - 2S^2 - i\ga^5 (\na_\al S^\al)
\label{P1}
\eeq
and
\beq
&&
\hat{\mathcal S}_{1,\,\al\be}
\,=\, \big[\na_\be ,\,\na_\al \big]
+ \na_\be {\hat h}_{1\al} - \na_\al {\hat h}_{1\be}
+  {\hat h}_{1\be} {\hat h}_{1\al}
 - {\hat h}_{1\al} {\hat h}_{1\be}
\nn
\\
&&
\qquad\quad
\,=\, - \,\frac14R_{\al\be\la\tau} \ga^\la \ga^\tau
- S^2 \big( \ga_\al \ga_\be  - \ga_\be \ga_\al \big)
- 2S^\la \big(S_\al \ga_\be  - S_\be \ga_\al \big)\ga_\la
\nn
\\
&&
\qquad\qquad
+\, \frac{i}{2}\ga^5
\Big[
\big(\na_\be S^\la\big)
\big( \ga_\la \ga_\al  - \ga_\al \ga_\la \big)
- \big(\na_\al S^\la\big)
\big( \ga_\la \ga_\be  - \ga_\be \ga_\la \big)\Big].
\label{S1}
\eeq

The general expression for the one-loop divergences is
\cite{DeWitt65}
\beq
\Ga^{(1)}_{div}
&=&
-\,\frac{\mu^{n-4}}{\vp}\,\int d^nx\sqrt{-g}\,
\tr
\bigg[
\frac{{\hat 1}}{180}\big(  R_{\mu\nu\al\be}^2
- R_{\al\be}^2
+ \cx R\big)
\nn
\\
&&
\quad
+ \,\,\frac12{\hat P}^2
+ \frac{1}{12}\hat{\mathcal S}_{\mu\nu}^2
+\frac16\,\cx {\hat P}\bigg],
\qquad
\qquad
\label{gendivs}
\eeq
where the trace and sign correspond to bosonic fields and
for the fermions the sign should be inverted. In the present case,
${\hat P}$ and $\hat{\mathcal S}_{\mu\nu}$ are defined by
(\ref{P1}) and (\ref{S1}). The calculation is pretty much
standard, but we quote simple relation for (\ref{CS}),
\beq
\frac12\, S_{\mu\nu}^2 \,=\,
(\na_\mu S_\nu)^2
- \big(\na_\mu S^\mu\big)^2
+ R_{\mu\nu} S^\mu S^\nu
+ 2 \na_\nu \big(S^\nu\na_\mu S^\mu - S^\mu \na_\mu S^\nu \big),
\label{transform}
\eeq
which proves useful, also, for integrating anomaly. Here
$(\na_\mu S_\nu)^2=\big(\na_\mu S_\nu\big)\big(\na^\mu S^\nu\big)$.

Finally, for the divergences we obtain the expression (with
recovered $\eta$)
\beq
&&
\Ga^{(1)}_{div,\,fer,\,1}
\,=\,
-\,\frac{\mu^{n-4}}{\vp}\,\int d^nx\sqrt{-g}\,\Big[
\frac{m^2}{3}R
+ 8m^2\eta^2S^2
- 2m^4
+ \frac{1}{20}C^2
- \frac{11}{360}E_4
+ \frac{1}{30} \cx R
\nn
\\
&&
\qquad\qquad
-\, \frac{2}{3} \eta^2 S_{\mu\nu}^2
+ \frac43 \eta^2 \cx S^2
- \frac43 \eta^2 \na_\be \big(S^\al \na_\al S^\be - S^\be \na_\al S^\al\big)
- \frac13\eta \,\na_\be B^\be\Big].
\qquad
\qquad
\label{ferdivs1}
\eeq
There are several remarkable aspects in this formula. In the
massless theory and in the limit $n\to 4$, the integrand is
conformal invariant, that is, composed by the $\mathcal{C}$-type and
$\mathcal{N}$-type invariants \cite{torsi}. In the expression for
divergences, one can identify three torsion-dependent total
derivatives. In our opinion, one of them is especially interesting,
albeit it turns out trivial (hence it was ignored in \cite{torsi}).
The last term in (\ref{ferdivs1}) depends on the vector field
\beq
B^\nu
= R^\nu_{\,\cdot\,\mu\tau\la}\vp^{\tau\la\al\mu}S_\al
= C^\nu_{\,\cdot\,\mu\tau\la}\vp^{\tau\la\al\mu}S_\al .
\label{Bdoido}
\eeq
Under the conformal transformation (\ref{confscal}), this vector
has a non-conventional transformation rule
$B^\nu = {\bar B}^\nu e^{-4\si}$ and, as a result,  $\na_\nu B^\nu$
is a curious example of the $\mathcal{C}$-type total derivative
invariant. Moreover, anticipating this part, the integration of the
corresponding total derivative term of the anomaly surprisingly
produces a nonlocal term in the effective action.
Unfortunately, the bright career of this term ends early, because
$B^\nu$ can be shown to vanish as a result of cyclic identity for a
Riemann (or Weyl) tensor. Anyway, it is remarkable that the general
rule of having nonlocal term in the action for a  $\mathcal{C}$-type
invariant in the anomaly holds in this case.

\subsection{Second calculation of fermion contribution}
\label{sec3.2}

The second scheme of doubling the fermion operator (\ref{H}) uses
the torsion-independent conjugate operator
\beq
{\hat H}_2
\,=\,i\ga^\nu \na_\nu + m.
\label{H2}
\eeq
In this case, one has to use the formula (\ref{product1}) for the
torsion-independent terms, which are certainly the same as in
(\ref{ferdivs1}). However, for the $S_\mu$-dependent terms,
$\Tr \ln {\hat H}_2$ is irrelevant and we have to use the modified
rule
\beq
- i \Tr \ln {\hat H}
\,=\,
- i \Tr \ln \big({\hat H} {\hat H}_2\big)
\,=\,
- i \Tr \ln
\big({\hat \cx} + 2{\hat h}_2^\al \na_\al + {\hat \Pi}_2\big).
\label{product2}
\eeq
The elements of the operator, in this case, are
\beq
&&
{\hat h}_2^\al \,=\, \frac{i}{2}\,\ga^5\ga^\la \ga^\al S_\la ,
\nn
\\
&&
{\hat \Pi}_2 \,=\, m^2 - \frac14R + m\ga^5 \ga^\la S_\la.
\label{hPi2}
\eeq

The elements of Schwinger-DeWitt technique are
also different,
\beq
&&
{\hat P}_2
\,=\,
m^2 - \frac{1}{12}R
- \frac12S^2
+  m\ga^5  \ga^\al S_\al
- \frac{i}{2}\ga^5 (\na_\al S^\al)
+ \frac{i}{4}\ga^5 \ga^\al \ga^\be S_{\al\be},
\label{P2}
\eeq
\beq
&&
\hat{\mathcal S}_{2,\,\al\be}
\,=\,
- \,\frac14R_{\al\be\la\tau} \ga^\la \ga^\tau
- \frac14 S^2 \big( \ga_\al \ga_\be  - \ga_\be \ga_\al \big)
- \frac12 S^\la \ga_\la \big(S_\be \ga_\al  - S_\al \ga_\be \big)
\nn
\\
&&
\qquad\qquad
+\, \frac{i}{2}\ga^5\ga^\la
\Big[\ga_\al \big(\na_\be S_\la\big)
- \ga_\be \big(\na_\al S_\la\big)
\Big].
\label{S2}
\eeq

Let us write only the $S_\mu$-dependent divergences, which are
obtained via (\ref{product2}) and (\ref{gendivs}),
\beq
&&
\Ga^{(1)}_{div,\,fer,\,2}
\,=\,
-\,\frac{\mu^{n-4}}{\vp}\,\int d^nx\sqrt{-g}\,\Big[
8m^2\eta^2 S^2
\, -\, \frac{2}{3}\eta^2 S_{\mu\nu}^2
\nn
\\
&&
\qquad\qquad\quad
+ \, \frac23 \eta^2 \cx S^2
+ \frac23 \eta^2\na_\be\big(S^\be \na_\al S^\al-S^\al\na_\al S^\be\big)
- \frac13\eta\,\na_\be B^\be\Big].
\qquad
\qquad
\label{ferdivs2}
\eeq

Compared to (\ref{ferdivs1}), the non-surface terms are the same.
However, the total derivative, $\mathcal{N}$-terms, have different
coefficients. This result represents the new kind of multiplicative
anomaly, being qualitatively different from the previously known
examples (starting from \cite{QED-Form}) concerning the
nonlocal part of the one-loop effective action. In these examples
the multiplicative anomaly shows up only for the massive fields and,
on the other hand, it cannot be compensated by the change of
renormalization condition because the last concerns only the
local terms. In the present case, the difference cannot be
compensated by the change of renormalization conditions for
the irrelevant surface integrals because such change given only
finite differences. As we will see in the next section, the finite
difference shows up in the local terms which are \textit{not}
total derivatives.

\subsection{Action of torsion and UV logarithmic corrections}
\label{sec3.3}

One of the important outputs of the one-loop calculations for
scalars and fermions is that, in the semiclassical conformal theory
with antisymmetric torsion, the classical action of torsion has the
form \cite{book,torsi}
\beq
S_{tors} \,=\,\int d^4x \sqrt{-g}
\Big\{
\, - a_1 S^4
-\, \frac{a_2}{4}\,  S_{\mu\nu}^2
+ b_1 \na_\be \big(S^\al \na_\al S^\be - S^\be \na_\al S^\al\big)
+ b_2 \cx S^2
\Big\},
\label{actSclas}
\eeq
where $a_{1,2}>0$ and  $b_{1,2}$ are arbitrary parameters. The
positiveness of $a_1$ and $a_2$ provides the tree-level potential
of $S_\mu$ bounded from below (as will be discussed in the 
next sections) and the positiveness of energy for propagating torsion
\cite{Bel-tors} (see also \cite{torsi}).

It may look natural to set
$a_2=1$   \cite{Bel-tors}, that can be provided by rescaling
$S_\mu$ and $\eta$. However, it is sometimes useful to keep
$a_2$ arbitrary, as we shall see in what follows. From the
viewpoint of conformal symmetry, $a_{1,2}$-structures
represent $\mathcal{C}$-terms and the values of those parameters can
be defined only from the measurements, which in the case of $a_2$
can be traded to the  measurement of $\eta S_\mu$. At the same
time, the coefficients of the $\mathcal{N}$-terms $b_{1,2}$ do not
affect equations of motion and are artificial parameters that cannot
be measured. Still, these terms are necessary for renormalizability
of a semiclassical theory.

Let us evaluate loop corrections to the vacuum action
(\ref{actSclas}). As a first step in this direction, we can
recover the leading one-loop logarithms in the most relevant
$\mathcal{C}$-terms. Using the standard considerations \cite{bavi85}
(see also \cite{OUP} for more details), we arrive at the one-loop
corrected torsion sector of the theory
\beq
\Ga^{(1)}_{\,tors}
\,=\, - \int d^4x \sqrt{-g}
\bigg\{
S^2\Big[a_1 + \frac{\be_1}{2} \ln \Big( \frac{\cx\,}{\mu^2}\Big) \Big] S^2
\,+\,
\frac{1}{4}\,S_{\mu\nu}\Big[a_2 + \frac{\be_2}{2}
\ln \Big( \frac{\cx\,}{\mu^2}\Big) \Big]\,S^{\mu\nu}
\Big\},
\label{Gammaln}
\eeq
From (\ref{scalardivs}) and (\ref{ferdivs1}) [see also
subsequent Eq.~(\ref{ferdivs2})], we can easily get
\beq
&&
\be_1
\,=\, -\,\frac{1}{2(4 \pi)^2}\,\sum_{i=1}^{N_s} \xi^2_{5,\,i} ,
\nn
\\
&&
\be_2
\,=\, \frac{8}{3(4 \pi)^2}\,\sum_{k=1}^{N_f} \eta^2_k .
\label{betas}
\eeq
Here $\xi_{5,\,i}$ and $\eta_k$ are nonminimal parameters for
different species of scalar and spinor fields. According to the
analysis of renormalization in interacting theories \cite{bush85},
these parameters may be different for different fields. Independent
on this, the signs of the beta functions show that the sign of
$\be_1$ indicated the asymptotic freedom in the parameter $a_1$
and the sign of $\be_2$ is positive, as it is typical for the
Abelian vector models. It is worth mentioning that these signs
correspond to the fermion and scalar contributions only, while
the contribution of the proper field $S_\mu$ was not taken into
account.

The integration of anomaly is, to a great extent, an elegant and
useful way to work with formula (\ref{Gammaln}) by constructing
a local version of renormalization group.
After deriving the covariant form of anomaly-induced action, we
use the duality of the UV and IR limits in the massless theory
and construct the low-energy alternative to (\ref{Gammaln}).

\section{Integration of anomaly with torsion}
\label{sec4}

Since the torsion field does not transform in (\ref{confscal}),
the derivation of anomaly has no novelties compared to the
purely metric case \cite{rie,frts84} (see, e.g. \cite{OUP} for
detailed introduction). On top of that, in \cite{anhesh} one can
find even more general consideration, with the torsion trace
$T_\mu$ included. Thus, we shall simply write down the
expression for the anomaly
\beq
&&
\langle T^\mu_{\,\,\mu} \rangle
\,=\,
-\, \Big\{wC^2 + bE_4 + c \cx R
-  \be_1 S^4
-  \frac14 \be_2 S_{\mu\nu}^2
\nn
\\
&&
\qquad \qquad
\,+\,
\ga_1 \na_\be \big(S^\al \na_\al S^\be - S^\be \na_\al S^\al\big)
\,+\, \ga_2 \cx S^2 \Big\}.
\label{anomaly}
\eeq
The one-loop $\be$-functions $\,w$, $b$ and $c\,$ do not depend
on the presence of torsion and are given by the expressions
\cite{birdav,OUP},
\beq
&&
\left(
\begin{array}{c}
w \\
b  \\
c  \end{array}
\right)
\,\,=\,\,
\frac{1}{360\,(4\pi)^2}\,
\left(
\begin{array}{c}
3N_s  + 18 N_f + 36 N_v   \\
- N_s  - 11\,N_f  - 62 N_v   \\
2 N_s + 12 N_f  - 36 N_v \end{array}
\right)\,.
\label{wbc}
\eeq
where $N_s$, $N_f$ and $N_v$ are the numbers of scalar, spinor
and gauge vector fields.

The beta functions $\be_{1,2}$ are written in (\ref{betas}).
Finally, the two functions $\ga_{1,2}$ in (\ref{anomaly}) are
ambiguous, as we have seen from the fermionic divergences
(\ref{ferdivs1}) and (\ref{ferdivs2}). For these two schemes
of calculation we meet, respectively,
\beq
&&
\ga_1^{(1)}
\,=\,- \frac{4}{3(4 \pi)^2}\,\sum_{k=1}^{N_f} \eta^2_k,
\qquad
\ga_1^{(2)}
\,=\, \frac{1}{3(4 \pi)^2}\,\sum_{k=1}^{N_f}\eta^2_k,
\label{gammas-1}
\\
&&
\ga_2^{(1)}
\,=\,\frac{4}{3(4 \pi)^2}\, \sum_{k=1}^{N_f} \eta^2_k
\,-\,
\frac{1}{6(4 \pi)^2}\,\sum_{i=1}^{N_s} \xi_{5,\,k},
\nn
\\
&&
\ga_2^{(2)}
\,=\,\frac{2}{3(4 \pi)^2}\,\sum_{k=1}^{N_f} \eta^2_k
\,-\, \frac{1}{6(4 \pi)^2}\,\sum_{i=1}^{N_s} \xi_{5,\,k}.
\label{gammas-2}
\eeq
Let us note that the scalar contributions to $\ga_2$ in
(\ref{gammas-2}), coming from (\ref{scalardivs}), also
have ambiguity, however one has to perform Pauli-Villars
analysis to see this. The required procedure would be a
mere repetition
of the one described in \cite{BoxAno} and \cite{AnoScal}
for background scalars, hence we skip this part.

In the rest of this section, we describe the solution of the equation
\beq
- \frac{2}{\sqrt{-g}}\,g_{\mu\nu}\,
\frac{\de \Ga_{ind}}{\de g_{\mu\nu}}
\,=\,
- \frac{1}{\sqrt{-\bar{g}}}\,e^{-4\si}\,
\frac{\de \Ga_{ind}}{\de \si}\bigg|
\,\,=\,\, \langle T^\mu_{\,\,\mu} \rangle .
\label{EA}
\eeq
The first equation here is an identity which uses $\si$, i.e., the
conformal factor of the metric defined in (\ref{confscal}).
Also, $\big| $ means the procedure of replacing
$\big(\bar{g}_{\mu\nu},\,\bar{S_\mu}\big)
\rightarrow \big( g_{\mu\nu},\,S_\mu\big)$ and $\si \rightarrow 0$.

The $4D$ solution for a purely gravitational case was found in
\cite{rie,frts84}. The generalization for a theory with torsion has been
found \cite{buodsh85,anhesh}, but only in the noncovariant
formulation as a functional of $\bar{g}_{\mu\nu}$, $\bar{S_\mu}$
and $\si$. In what follows, we shall construct the most informative,
covariant (nonlocal and local) solutions following the general scheme
working for an arbitrary even dimension \cite{6d}. Thus, we need
just to give a practical realization of this scheme for the theory
with torsion.

The conformal invariants in (\ref{anomaly}) can be denoted in
a common way as
\beq
&&
Y
\,=\,
Y(g,S)
\,=\,
wC^2 - \be_1 S^4 -  \frac14 \be_2 S_{\mu\nu}^2.
\label{Y}
\eeq
The unique topological term $E_4$ has the remarkable conformal
property
\beq
\sqrt{-g}\,\Big(E_4-\frac23\,{\cx} R\Big)
\,=\,
\sqrt{-\bar{g}}\,\Big({\bar E_4}-\frac23\,{\bar \cx} {\bar R}
+ 4{\bar \De_4}\si \Big),
\label{119}
\eeq
where $\Delta_4 =
\cx^2 + 2R^{\mu\nu}\nabla_{\mu}\nabla_{\nu} - \dfrac{2}{3}R\square
+\dfrac{1}{3}(\nabla^{\mu}R)\nabla_{\mu}$, which obeys
$\sqrt{-g}\Delta_4=\sqrt{-\bar{g}}\bar{\Delta}_4$
\cite{FrTs-superconf,Paneitz}.

These notations and features do not depend on the presence of torsion
and, therefore, we can directly write down the nonlocal part of the
solution of (\ref{EA}),
\beq
&&
\Ga_{ind,\,nonloc}\,=\,
\frac{b}{8}\int_x\int_y\Big(E_4
-\frac23\square R\Big)_{\hspace{-1mm}x}
G(x,y)\Big(E_4-\frac23\square R\Big)_{\hspace{-1mm}y}
\nn
\\
&&
\qquad \quad
+\,
\frac{1}{4}\int_x\int_y Y(x)\, G(x,y)
\Big(E_4-\frac23\square R\Big)_{\hspace{-1mm}y},
\label{nonlocal-S}
\eeq
where we used the notation $\int_x\equiv \int d^4 x\sqrt{-g(x)}$
and the Green function of the Paneitz operator
\beq
(\sqrt{-g}\De_4)_xG(x,y)\,=\,\de(x,y).
\label{Green_function}
\eeq

Let us find a solution for the
total derivative terms. For the $\cx R$ the result is well-known,
\beq
&&
-\frac{2}{\sqrt{-g}}g_{\mu\nu}\frac{\delta}{\delta g_{\mu\nu}}
\int_x R^2\,=\,12\square R.
\label{TrR2}
\eeq
and for the $\cx S^2$ the answer can be easily found by direct
calculation,
\beq
&&
-\,\frac{2}{\sqrt{-g}}\,g_{\mu\nu}\,\frac{\de}{\de g_{\mu\nu}}
\int_x RS^2\,=\,6\square S^2.
\label{RS2}
\eeq

Thus, the remaining problem is to integrate the $\ga_1$ term in
(\ref{anomaly}). Let us use the hypothesis that, as in all previously
known cases, the solution for the total derivative should be a local
covariant action. Then we have the following candidate terms:
\beq
\Ga_{ind,\,local} \,=\,
\int_x \big\{
\al_1  (\na_\mu S^\mu)^2
\,+\,
\al_2 (\na_\mu S_\nu)^2
\,+\,
\al_3  R S^2 \big\},
\label{localS}
\eeq
where the last one is already worked out in (\ref{RS2}). We can
rewrite the \textit{r.h.s.} of this formula using
$\cx S^2 = 2\na_\nu (S_\mu \na^\nu S^\mu)$.  It is easy to note that
in (\ref{localS}) the dimensionally possible term
$R_{\mu\nu} S^\mu S^\nu$ is missing. The reason is that the
linear combination (\ref{transform}) gives conformal invariant
functional $\int_x S_{\mu\nu}^2$ and, therefore, including the
mentioned term would be senseless. The application of the
conformal operator to the remaining two terms gives
\beq
&&
-\frac{2}{\sqrt{-g}}g_{\mu\nu}\frac{\de}{\de g_{\mu\nu}}
\int_x (\na_\mu S^\mu )^2
\,=\,
4 \na_\nu (S^\nu \na_\mu S^\mu),
\nn
\\
&&
-\frac{2}{\sqrt{-g}}g_{\mu\nu}\frac{\de}{\de g_{\mu\nu}}
\int_x (\na_\nu S_\mu )^2
\,=\,
2 \na_\nu \big[
  S^\nu \na_\mu S^\mu
- S_\mu \na^\nu S^\mu
- S_\mu \na^\mu S^\nu
\big].
\label{alphas12}
\eeq

Using (\ref{alphas12}) together with the modified version of
(\ref{RS2}), replacing the result into the linear combination of
(\ref{localS}) and comparing to (\ref{anomaly}), we arrive at
the solution for $\al_{1,2,3}$
\beq
\al_1 = 0,
\qquad
\al_2 = \frac12\,\ga_1,
\qquad
\al_3 = \frac{1}{12}(\ga_1 - 2\ga_2).
\label{alphs}
\eeq
Taking into account relations (\ref{TrR2}) and (\ref{119}),
the local part of the covariant induced action has the form
\beq
&&
\Ga_{ind,\,loc}
\,=\,
- \,\frac{3c+2b}{36}\int_x R^2
\,\,+\, \int_x \Big\{\frac{\ga_1}{2}(\na_\mu S_\nu )^2
\,+\,
\frac{\ga_1 - 2\ga_2}{12}\,RS^2 \Big\}.
\label{locEA}
\eeq
Just to complete the story, we mention that this expression may
be modified by using the relations (\ref{Weyl-1}) and (\ref{GB})
in the purely metric part and (\ref{transform}) in the torsion-metric
part. This means, one can use the replacement
$R^2 \to \frac13 R^2_{\mu\nu}$ or
$R^2 \to \frac13 R^2_{\mu\nu\al\be}$ in (\ref{locEA}) and use
(\ref{transform}) to make similar trades in the $S$-dependent terms.

It is worth mentioning another detail concerning fermion
contributions. The local torsion-dependent terms (\ref{locEA})
violate not only conformal (\ref{confscal}), but also chiral
symmetry (\ref{chiral}). This symmetry breaking does not occur
in the fermionic non-local part (\ref{nonlocal-S}).

All in all, the general covariant solution for the anomaly-induced
action is the sum of the nonlocal (\ref{nonlocal-S}) and local
(\ref{locEA}) parts,
\beq
&&
\Ga_{ind} \,=\, S_c(g,S) \,+\, \Ga_{ind,\,nonloc}
\,+\, \Ga_{ind,\,loc}\,,
\label{nonlocal}
\eeq
where $S_c(g,S)$ is an arbitrary conformal invariant functional
which plays the role of integration constant for our main equation
(\ref{EA}). The uncertain elements in this expression are this
unknown functional and the ambiguous $\gamma$-functions in
the local part $\Ga_{ind,\,nonloc}$ in (\ref{locEA}). Similarly to
the ambiguity in the $R^2$-term, these torsion-dependent local terms
may be modified by adding the local non-conformal terms to the
\textit{classical} action of vacuum (\ref{actSclas}). These
classical terms are not subject of renormalization and represent
a new type of arbitrariness in the action, equivalent to the local
multiplicative anomaly.

As usual, we can rewrite the nonlocal part of (\ref{nonlocal}) in the
symmetric form and get the induced action in the local covariant
form with two auxiliary fields $\varphi$ and $\psi$ \cite{a}
(see also \cite{MazMott01}),
\beq
&&
\Gamma_{ind}
\,=\,
S_c(g,S) \,+\, \Ga_{ind,\,loc}
\,+\,
\int_{x}\bigg\{\frac12 \ph \Delta_4\ph - \frac12 \psi\De_4\psi
\nn
\\
&&
\qquad\quad
+\,\,
\frac{\sqrt{-b}}{2}\varphi
\Big(E_4-\frac23\square R+\frac{1}{b}Y\Big)
\,+\, \frac{1}{2\sqrt{-b}}\,\psi Y
\bigg\},
\label{aux_fields}
\eeq
where the local part and $Y$ are given by (\ref{locEA}) and
(\ref{Y}), respectively.

The forms  \eqref{nonlocal} and \eqref{aux_fields} are equivalent
to the noncovariant form derived in \cite{buodsh85}. Each of this
forms has its own advantages, in particular (\ref{aux_fields}) is
more suitable for physical applications \cite{balsan,PoImpo}. On
another hand, the nonlocal form (\ref{nonlocal}) is more explicit
and, also, was recently shown to admit the description of the IR
limit \cite{MotVaul,AnoScal}. We shall apply this approach to the
induced action with torsion (\ref{nonlocal}) in the next section.

\section{Anomaly-induced effective action in the IR}
\label{sec5}

In the recent works \cite{BTSh} and \cite{Shap-tor} it was shown that
dynamical torsion may be used to construct phenomenologically
successful models of dark matter (DM). On the other hand, there is
a general statement that the
consistency of quantum field theory of the propagating torsion
requires a large torsion mass \cite{Bel-tors,guhesh}, something that
can be in contradiction to the DM applications. In this respect, it
looks interesting to explore the possibility of dynamical symmetry
breaking in the torsion sector. In scalar field theory, this is one
of the ways have a large mass in the IR and, at the same time,
leave some space for the applications in the high energy physics,
including to early Universe.

In the scalar case, the analysis of symmetry breaking in initially
massless theory requires the effective potential \cite{ColeWein},
that can be also derived in curved spacetime \cite{book,OUP}.
We shall follow \cite{AnoScal}, where the scalar potential was
obtained in the IR limit of the anomaly-induced action,
i.e., the scalar analog of \eqref{nonlocal}. We shall concentrate
only on the nonlocal part of this action because the local part is
ambiguous.

Let us define the meaning of the low-energy (IR) limit in the
massless conformal theory, with $\xi_1= 1/6$. The main assumption
is that torsion terms in (\ref{Y}) dominate over the square of the
Weyl tensor. This may be a reasonable approximation in the early
Universe because Weyl tensor vanishes for the homogeneous and
isotropic metric and shows up only because of the metric
perturbations. On the other hand, one can assume that torsion plays
an important role in the formation of DM and hence should be a
strong field \cite{Shap-tor}. Thus,
\beq
\big| S^4\big|  \gg \big|C_{\mu\nu\al\be}^2\big|
\qquad
\mbox{and}
\qquad
\big| S_{\mu\nu}^2\big|  \gg \big|C_{\mu\nu\al\be}^2\big|.
\label{IR-1}
\eeq
As usual in general relativity, the IR limit implies a weak
gravitational field. The weak gravity can be described by a small
metric perturbation $h_{\mu\nu}=g_{\mu\nu}-\eta_{\mu\nu}$, that
means, e.g.,  $|\cx R| \gg |R^2_{\ldotp\ldotp\ldotp\ldotp}|$ for all
curvature tensors (e.g., Weyl, Ricci tensors, and $R$).

In this approximation, the Green function (\ref{Green_function})
reduces to
\beq
G
\,=\,\Delta^{-1}_{4}
\,=\,\Big(\square^2
+ 2R^{\mu\nu}\nabla_{\mu}\nabla_{\nu}
- \dfrac{2}{3}R\square
+ \dfrac{1}{3}R^{;\mu} \nabla_{\mu}\Big)^{-1}
\,\approx \,\frac{1}{\square^2}.
\label{apprG}
\eeq
Thus, the nonlocal, torsion-dependent part of
the effective action (\ref{nonlocal-S}) boils down to
\beq
\Ga^{IR}_{ind,\,nonloc}&=&
\frac{1}{6}
\int_x\int_y \Big(\be_1 S^4 +  \frac14 \be_2 S_{\mu\nu}^2\Big)_x
\Big( \frac{1}{\cx^2}\Big)_{x,y}
\,\Big(\cx R\Big)_{\hspace{-1mm}y}
\nn
\\
&&
=\,\,
\frac{1}{6}
\int_x\int_y \Big(\be_1 S^4 +  \frac14 \be_2 S_{\mu\nu}^2\Big)_x
\Big( \frac{1}{\cx}\Big)_{x,y}\,R(y).
\label{nonlocal-IR}
\eeq
On top of this expression, the IR limit of the induced effective
action includes $\mathcal{O}(R^ 2_{...})$-terms, but those were
discussed in \cite{AnoScal} and we can refer the interested reader
to this work.

In the presence of torsion, the terms $S^4 \cx^{-1}R$ and
$S_{\mu\nu}^2 \cx^{-1}R$ have the same \textit{global}
scaling as the respective classical terms $S^4$ and
$S_{\mu\nu}^2$, i.e., they are invariant
under the transformation (\ref{confscal}) with $\si \to \la = const$.
Indeed, this is the usual feature of the nonlocal induced action,
independent on extra fields and even spacetime dimension
\cite{6d}, but it is quite remarkable that this feature holds in the
IR limit, just as in the scalar case \cite{AnoScal}.

The next step is to derive the low-energy effective action of
torsion from (\ref{nonlocal-IR}). To this end, we separate the
conformal factor of the metric and use the analogy with the
renormalization group - based derivation of effective action
\cite{BuchWolf,book}. At one loop, it is sufficient to account
only for the linear in $\si$ terms. Thus, we consider
\beq
&&
g_{\mu\nu}=\bar{g}_{\mu\nu}e^{2\sigma},
\qquad
\sqrt{-g}S^4 = \sqrt{-\bar{g}}\bar{S}^4,
\qquad
\sqrt{-g}S_{\mu\nu}^2 = \sqrt{-\bar{g}}\bar{S}_{\mu\nu}^2,
\nn
\\
&&
{\cx}^{-1} \,=\,e^{2\sigma}\bar{\cx}^{-1},
\qquad
R\,=\,e^{-2\si}\big[\bar{R}-6\bar{\cx}\sigma\big],
\label{R-exp}
\eeq
where $\bar{\cx}=\bar{g}^{\mu\nu}\partial_\mu\partial_\nu$.
In this framework, (\ref{nonlocal-IR}) becomes
\beq
\Ga^{IR}_{ind,\,nonloc}&=&
\int_x \Big( \be_1 \bar{S}^4
- \frac14 \be_2 \bar{S}_{\mu\nu}^2 \Big)_x \si(x).
\label{nonlocal-IR-si}
\eeq
This result demonstrates, as we expected, that the anomaly-induced
action is a local version of the renormalization group corrected
classical action (\ref{actSclas}), that means the substitution
\beq
a_1 \longrightarrow a_1 + \be_1 \si(x),
\qquad
a_2 \longrightarrow a_2 + \be_2 \si(x).
\label{RGimprove}
\eeq
Compared to the usual curved-space renormalization group
\cite{tmf,OUP}, the constant scaling parameter $\la$
is traded for the coordinate-dependent conformal factor of the
metric $\si$, i.e., we arrive at the local form of renormalization
group in curved space \cite{PoImpo}.

At this point, one can use (\ref{RGimprove}) to recover the
low-energy effective action. This requires identification of the
scale parameter $\si$ and we have several choices because of
the scaling rules
\beq
S^2 = \bar{S}^2 e^{-2\si},
\qquad
S^4 = \bar{S}^4 e^{-4\si},
\qquad
S_{\mu\nu}^2 = \bar{S}_{\mu\nu}^2 e^{-4\si}.
\label{scale12}
\eeq
E.g., choosing the first option, we arrive at the identification
$\sigma \to \frac12 \ln \frac{S^2}{\mu^2}$. Then, the
improvement (\ref{RGimprove}) of the action (\ref{actSclas})
gives, in the torsion-dependent sector,
\beq
\Ga_{tors} \,=\,- \int d^4x \sqrt{-g}
\bigg\{
\Big[a_1 +  \frac{\be_1}{2} \ln \Big(\frac{S^2}{\mu^2}\Big)\Big]S^4
\,+\, \frac14 \Big[a_2
+  \frac{\be_2}{2} \ln \Big(\frac{S^2}{\mu^2}\Big)\Big] S_{\mu\nu}^2
\,+ \,...\bigg\},
\label{IR-Simp}
\eeq
where we omitted surface terms.
An obvious observation here is that (\ref{IR-Simp}) is not just an
integral of the effective potential since there is a kinetic term
$S_{\mu\nu}^2$. Thus, the result can be seen as a form of the
one-loop effective action in the IR limit.

The effective potential part of (\ref{IR-Simp}) has the form
\beq
V_{eff}
\,=\,
\Big[a_1 +  \frac{\be_1}{2} \ln \Big(\frac{S^2}{\mu^2}\Big)\Big]S^4,
\label{VS2}
\eeq
together with the negative $\beta_1$-function (\ref{betas}) shows
that the one-loop potential always becomes unstable for  large
values of $S^2$, where the quantum corrections start to dominate
over the classical coefficient $a_1$. The coefficients $\eta$ for
fermions are experimentally bounded by very small values, at least
for electrons (one can use \cite{torsi} as a starting point for further
references on the subject). Thus, according to  (\ref{betas}), the
strong effect of the negative $\beta_1$ may be expected only for
extremely large values of $S^2$. Anyway, at the one-loop level
the effective potential is unbounded from below.

\begin{figure}
\begin{quotation}
 \mbox{\hspace{+2.9cm}}
\includegraphics[width=6.8cm,angle=0]{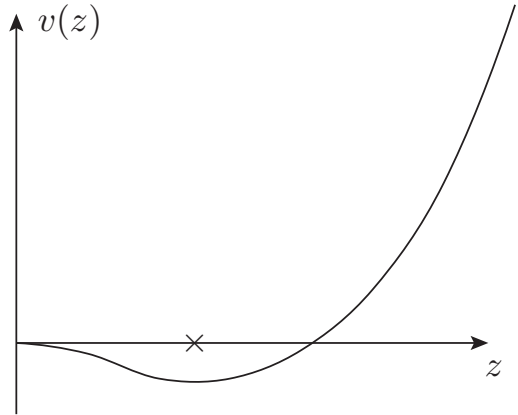}
\caption{{\sl
Plot of $v(z)$ demonstrating the possibility of dynamical
symmetry breaking for a positive $\be_1$.}}
\label{Fig1}
\end{quotation}
\end{figure}

This feature does not mean that the theory, in general, is badly
defined at the quantum level because the second and higher loop
contributions may restore the positive definiteness of the potential.
On the other hand, assuming the change of sign of $\be_1$ at
higher loops, we can rewrite the effective potential part of
(\ref{IR-Simp}) in terms of the dimensionless parameter
$z=S^2/\mu^2$
\beq
V_{eff}
\,=\,
\mu^4 v(z)
\,=\,
a_1 \mu^4 \big[ z^2 \big(1 + \tilde{\be}\ln z \big)\big],
\qquad
\tilde{\be} = \frac{\be_1}{2 a_1}\,.
\label{vz}
\eeq
The qualitative profile of the function $v(z)$ for $\be_1 > 0$ is
shown in Fig.~\ref{Fig1}. However, since the real sign of the
beta function is negative, the implementation of the dynamical
symmetry breaking in this theory requires further investigation
and, especially, higher loops contributions to the potential.

\section{Conclusions and discussions}
\label{sec6}

We calculated the vacuum divergences and formulated, for the
first time, the covariant version of the anomaly-induced effective
action in curved spacetime with torsion. The output can be presented
in the covariant nonlocal form (\ref{nonlocal}) or in the local form
with auxiliary scalars  (\ref{aux_fields}). The main novelty is
the detection of the multiplicative anomaly in the total derivative
part of the divergences, i.e., (\ref{ferdivs1}) vs  (\ref{ferdivs2})
and the corresponding ambiguity
in the local part of the induced effective action. The ambiguity
cannot be removed by the change of renormalization condition and
represents a new feature of massless fermionic determinants that
does not take place without torsion.

Multiplicative anomaly appears in the finite local part of induced
effective action, as shown in expression (\ref{locEA}). All the terms
in this action are ambiguous. The coefficient $c$ may be modified
by adding the finite $\int_x R^2$ term to the classical action or,
equivalently, by the choice of the divergent Weyl-squared
counterterm \cite{anomaly-2004}. On the
other hand, the ambiguity in the torsion-dependent terms can be
compensated by adding the local nonconformal terms similar to
those in (\ref{locEA}), to the classical action (\ref{actSclas}).

Another new result of our work is the covariant expression for
the low-energy limit of the anomaly-induced effective action
(\ref{IR-Simp}). This part may be eventually useful for describing
dynamical symmetry breaking in torsion theories but, independent
on that, we have an interesting analogy with the scalar effective
potential in the axial vector model. On the other hand,
the low-energy effective action (\ref{IR-Simp}) by itself may
serve as an evidence of breaking local conformal symmetry
by quantum corrections. In this sense, it is an analog of the
effective potential of a scalar field $\phi$ in the conformal theory,
where the $\phi^4 \ln(\phi/\mu)$-term breaks the symmetry of the
classical $\phi^4$-type potential. It looks remarkable that we can
obtain this low-energy breaking with torsion from the
anomaly-induced effective action (\ref{nonlocal}).

\section*{Acknowledgements}

G.C. is grateful to CAPES for supporting their Ph.D. project.
I.Sh. is partially supported by Conselho Nacional de Desenvolvimento
Cient\'{i}fico e Tecnol\'{o}gico - CNPq (Brazil) under the grant
303635/2018-5; by Funda\c{c}\~{a}o de Amparo \`a Pesquisa de
Minas Gerais - FAPEMIG under the project PPM-00604-18;
and by the Ministry of Education of the Russian Federation, under
the project No. FEWF-2020-0003.



\begin{thebibliography}{99}
		
\small{

\bibitem{CapDuf-74} D.M. Capper, M.J. Duff and L. Halpern,
\textit{Photon corrections to the graviton propagator,}
Phys. Rev. {\bf D10} (1974) 461;
\\
D.M. Capper and M.J. Duff,
\textit{The one-loop neutrino contribution 
to the graviton propagator,}
Nucl. Phys. {\bf B82} (1974) 147.

\bibitem{duff77} M.J. Duff,
\textit{ Observations On Conformal Anomalies,}
Nucl. Phys. {\bf B125} (1977) 334.
		
\bibitem{ddi} S. Deser, M.J. Duff and C. Isham,
\textit{ Nonlocal conformal anomalies},
Nucl. Phys. {\bf B111} (1976) 45.

\bibitem{duff94} M.J. Duff,
\textit{ Twenty years of the Weyl anomaly,}
Class. Quant. Grav. {\bf 11} (1994) 1387,
hep-th/9308075.

\bibitem{polyakov81} A.M. Polyakov,
\textit{ Quantum geometry of bosonic strings,}
Phys. Lett. {\bf B207} (1981) 211.

\bibitem{rie} R.J. Riegert,
\textit{ A non-local action for the trace anomaly},
Phys. Lett. {\bf B134} (1984) 56. 
		
\bibitem{frts84} E.S. Fradkin and A.A. Tseytlin,
\textit{ Conformal anomaly in Weyl theory and anomaly free
superconformal theories,}
Phys. Lett. {\bf B134} (1984) 187.

\bibitem{ChrFull} S.M. Christensen and S.A. Fulling,
\textit{Trace anomalies and the Hawking effect,}
\textit{ Phys. Rev.} {\bf D15} (1977) 2088.  

\bibitem{balsan}
R. Balbinot, A. Fabbri and I.L. Shapiro,
\textit{Anomaly induced effective actions and Hawking radiation,}
Phys. Rev. Lett. {\bf 83} (1999) 1494, 
hep-th/9904074;
\textit{Vacuum polarization in Schwarzschild space-time
by anomaly induced effective actions and Hawking radiation,}
Nucl. Phys. {\bf B559} (1999) 301,  
hep-th/9904162.

\bibitem{fhh} M.V. Fischetti, J.B. Hartle and B.L. Hu,
\textit{Quantum effects in the early Universe. 1. Influence of trace
anomalies on homogeneous, isotropic, classical geometries,}
Phys. Rev. {\bf D20} (1979) 1757.

\bibitem{star} A.A. Starobinsky,
\textit{A new type of isotropic cosmological models without
singularity},
Phys. Lett. {\bf B91} (1980) 99.

\bibitem{StabInstab}
T.d.P.~Netto, A.M.~Pelinson, I.L.~Shapiro and A.A.~Starobinsky,
\textit{ From stable to unstable anomaly-induced inflation,}
Eur. Phys. J. {\bf C76} (2016)  544,
arXiv:1509.08882.

\bibitem{PoImpo} I.L.~Shapiro,
\textit{ Effective action of vacuum: semiclassical approach},
Class. Quant. Grav. {\bf 25} (2008) 103001,
arXiv:0801.0216.

\bibitem{OUP} I.L. Buchbinder and I.L. Shapiro,
\textit{Introduction to Quantum Field Theory with Applications
to Quantum Gravity} (Oxford University Press, 2021).

\bibitem{Mottola-SM} E.~Mottola,
\textit{The Effective theory of gravity and dynamical vacuum energy,}
arXiv:2205.04703.

\bibitem{odsh90-CQG} S.D.~Odintsov and I.L.~Shapiro,
\textit{Perturbative approach to induced quantum gravity,}
Class. Quant. Grav. \textbf{8} (1991) L57. 

\bibitem{AntMot92} I.~Antoniadis and E.~Mottola,
\textit{$4-D$ quantum gravity in the conformal sector,}
Phys. Rev. \textbf{D45} (1992) 2013. 

\bibitem{induce} I.L. Shapiro,
\textit{Hilbert-Einstein action from induced gravity coupled
with scalar field,}
Mod. Phys. Lett. \textbf{A9} (1994) 1985;  
I.L. Shapiro and G. Cognola,
\textit{Interaction of low - energy induced gravity with quantized
matter and phase transition induced by curvature,}
Phys. Rev. {\bf D51} (1995) 2775, 
hep-th/9406027.

\bibitem{KamSch-2011} Z. Komargodski, and A. Schwimmer,
\textit{On Renormalization Group Flows in Four Dimensions,}
JHEP {\bf 1112} (2011) 099,
arXiv:1107.3987.

\bibitem{LPR-2012} M.A. Luty, J. Polchinski, and R. Rattazzi,
\textit{The a-theorem and the asymptotics of 4D quantum field theory,}
JHEP {\bf 1301} (2013) 152,
arXiv:1204.5221.

\bibitem{Kamarg-2012} Z. Komargodski,
\textit{The Constraints of conformal symmetry on RG flows,}
JHEP {\bf 1207} (2012) 069,
\\
arXiv:1112.4538.

\bibitem{nelspan82} B.L. Nelson and P. Panangaden,
\textit{Scaling behavior of interacting quantum fields in
curved space-time,}
Phys.Rev. {\bf D25} (1982) 1019.

\bibitem{tmf} I.L. Buchbinder,
\textit{ On Renormalization group equations in curved space-time,}
Theor. Math. Phys. {\bf 61} (1984) 393.

\bibitem{Toms83} D.J. Toms,
\textit{ The effective action and the renormalization group
equation in curved space-time,}
Phys. Lett. \textbf{B126} (1983) 37.

\bibitem{MotVaul} E. Mottola and R. Vaulin,
\textit{ Macroscopic effects of the quantum trace anomaly,}
Phys. Rev. {\bf D74} (2006) 064004,
gr-qc/0604051;
\\		
M. Giannotti and E. Mottola,
\textit{ Trace anomaly and massless scalar degrees of freedom in gravity,}
Phys. Rev. {\bf D79} (2009) 045014,
arXiv:0812.0351.

\bibitem{AnoScal} M.~Asorey, W.C.~Silva, I.L.~Shapiro and
P.R.B.~do Vale,
\textit{Trace anomaly and induced action for a metric-scalar
background,}
arXiv:2202.00154. 

\bibitem{DeserSchwimmer}
S.~Deser and A.~Schwimmer,
\textit{ Geometric classification of conformal anomalies in
arbitrary dimensions,}
Phys. Lett.  \textbf{B309} (1993) 279, 
hep-th/9302047.

\bibitem{6d} F.M. Ferreira and I.L. Shapiro,
\textit{ Integration of trace anomaly in $6D$,}
Phys. Lett. {\bf B772} (2017) 174, 
arXiv:1702.06892.

\bibitem{FST} F.M. Ferreira, I.L. Shapiro and P.M. Teixeira,
\textit{ On the conformal properties of topological terms in even
dimensions}
Eur. Phys. J. Plus {\bf 131} (2016) 164,
arXiv:1507.03620.

\bibitem{BFT-2000} 	
F. Bastianelli, S. Frolov, and A.A. Tseytlin,
\textit{ Conformal anomaly of (2,0) tensor multiplet in six-dimensions
and AdS / CFT correspondence,}
JHEP {\bf 02} (2000) 013,
hep-th/0001041;
\\
R.R. Metsaev,
\textit{6d conformal gravity,}
Journ. Phys. {\bf A44} (2011) 175402,
arXiv:1012.2079;
\\
H. L$\ddot{\rm u}$, Yi Pang, C.N. Pope,
\textit{Conformal gravity and extensions of critical gravity,}
Phys. Rev. {\bf D84} (2011) 064001,
arXiv:1106.4657.

\bibitem{anomaly-2004} M. Asorey, E.V. Gorbar and I.L. Shapiro,
\textit{Universality and ambiguities of the conformal anomaly,}
Class. Quant. Grav. {\bf 21} (2004) 163,
hep-th/0307187.

\bibitem{BoxAno}
M. Asorey, G. de Berredo-Peixoto and I.L. Shapiro,
\textit{Renormalization ambiguities and conformal anomaly
in metric-scalar backgrounds,}
Phys. Rev. {\bf D74} (2006) 124011,
arXiv:hep-th/0609138.

\bibitem{Obukhov83}
Y.N.~Obukhov,
\textit{Spectral geometry of the Riemann-Cartan space-time,}
Nucl. Phys.  \textbf{B212} (1983) 237. 

\bibitem{buodsh85}
I.L.~Buchbinder, S.D.~Odintsov and I.L.~Shapiro,
\textit{Nonsingular cosmological model with torsion induced by
vacuum quantum effects,}
Phys. Lett.  \textbf{B162} (1985) 92. 

\bibitem{anhesh} J.A. Helayel-Neto, A. Penna-Firme and I.L. Shapiro,
\textit{ Conformal symmetry, anomaly and effective action for
metric-scalar gravity with torsion,}
Phys. Lett. {\bf B479} (2000) 411,
gr-qc/9907081.

\bibitem{torsi} I.L.~Shapiro,
\textit{Physical aspects of the space-time torsion,}
Phys. Rept. \textbf{357} (2002) 113,
hep-th/0103093.

\bibitem{QED-Form}
B. Goncalves, G. de Berredo-Peixoto and I.L. Shapiro,
\textit{One-loop corrections to the photon propagator in the
curved-space QED,}
Phys.Rev. \textbf{D80} (2009) 104013,
arXiv:0906.3837

\bibitem{Evans} T.S. Evans,
\textit{Regularization schemes and the multiplicative anomaly,}
Phys. Lett. \textbf{B457} (1999) 127,
hep-th/9803184.

\bibitem{Dowker} J.S. Dowker,
\textit{ On the relevance of the multiplicative anomaly},
hep-th/9803200.

\bibitem{zeta} E. Elizalde,
\textit{ Zeta regularization techniques with applications},
(World Scientific, 1994).

\bibitem{Hehl-76}
F.W. Hehl, P. Heide, G.D. Kerlick and J.M. Nester,
\textit{ General relativity with spin and torsion: foundations and
prospects,} Rev. Mod. Phys.
{\bf 48} (1976) 393.

\bibitem{bush90} I.L.~Buchbinder and I.L.~Shapiro,
\textit{On the renormalization group equations in curved space-time
with torsion,}
Class. Quant. Grav. \textbf{7} (1990) 1197. 

\bibitem{bush85} I.L. Buchbinder and I.L. Shapiro,
\textit{On the renormalization of the models of quantum field
theory in an external gravitational field with torsion,}
Phys. Lett. {\bf B151} (1985)  263.

\bibitem{birdav} N.D. Birell and P.C.W. Davies,
\textit{Quantum fields in curved space,}
(Cambridge University Press, Cambridge, 1982).

\bibitem{DeWitt65} B.S. DeWitt,
\textit{ Dynamical theory of groups and fields,}
(Gordon and Breach, 1965).

\bibitem{bavi85}  A.O. Barvinsky and G.A. Vilkovisky,
\textit{ The generalized Schwinger-DeWitt technique in gauge theories
and quantum gravity,}
Phys. Repts. {\bf 119} (1985) 1.

\bibitem{GBP} G.B. Peixoto,
\textit{ A note on the heat kernel method applied to fermions,}
Mod. Phys. Lett. {\bf A16} (2001) 2463, 
hep-th/0108223.

\bibitem{FrTs-superconf} E.S. Fradkin and A.A. Tseytlin,
\textit{ Asymptotic freedom on extended conformal supergravities,}
Phys. Lett. {\bf B110} (1982) 117; 
\textit{ One-loop beta function in conformal supergravities,}
Nucl. Phys. {\bf B203} (1982) 157. 
		
\bibitem{Paneitz} S. Paneitz,
\textit{ A quartic conformally covariant differential operator for
arbitrary pseudo Riemannian manifolds,}
MIT preprint - 1983; SIGMA {\bf 4} (2008) 036,
arXiv:0803.4331.

\bibitem{a} I.L. Shapiro and A.G. Jacksenaev,
\textit{ Gauge dependence in higher derivative quantum gravity and
the conformal anomaly problem,}
Phys. Lett. {\bf B324} (1994) 286.
		
\bibitem{MazMott01} P.O. Mazur and E. Mottola,
\textit{ Weyl cohomology and the effective action for conformal
anomalies,}
Phys. Rev. {\bf D64} (2001) 104022,
hep-th/0106151.
		

\bibitem{BTSh}
A.S.~Belyaev, M.C.~Thomas and I.L.~Shapiro,
\textit{Torsion as a Dark Matter Candidate from the Higgs Portal,}
Phys. Rev. \textbf{B95} (2017) 095033,
arXiv:1611.03651.

\bibitem{Shap-tor}
M.~Shaposhnikov, A.~Shkerin, I.~Timiryasov and S.~Zell,
\textit{Einstein-Cartan Portal to Dark Matter,}
Phys. Rev. Lett. \textbf{126} (2021) 161301;
Erratum: ibid, \textbf{127} (2021)  169901,
arXiv:2008.11686.

\bibitem{guhesh}
G.~de Berredo-Peixoto, J.A.~Helayel-Neto and I.L.~Shapiro,
\textit{On the consistency of a fermion torsion effective theory,}
JHEP \textbf{02} (2000) 003,
hep-th/9910168.

\bibitem{Bel-tors} A.S.~Belyaev and I.L.~Shapiro,
\textit{The Action for the (propagating) torsion and the limits on
the torsion parameters from present experimental data,}
Phys. Lett. \textbf{B425} (1998) 246,  
hep-ph/9712503.

\bibitem{book} I.L. Buchbinder, S.D. Odintsov and I.L. Shapiro,
\textit{Effective Action in Quantum Gravity}
(IOP Publishing, Bristol, 1992).

\bibitem{ColeWein} S.R. Coleman and E.J. Weinberg,
\textit{ Radiative corrections as the origin of spontaneous symmetry
breaking,} Phys. Rev. {\bf D7} (1973) 1888.

\bibitem{BuchWolf} I.L. Buchbinder and J.J. Wolfengaut,
\textit{Renormalization group equations and effective action in
curved space-time,}
Class. Quant. Grav. {\bf 5} (1988) 1127. 
}

\end{thebibliography}
	
\end{document}